\documentclass[12pt]{amsart}
\newtheorem{theorem}{Theorem}[section]

\newtheorem{lemma}{Lemma}[section]
\newtheorem{prop}{Proposition}[section]
\newtheorem{coro}{Corollary}[section]

\newtheorem{exampleHLP}{Example}[section]
\newenvironment{example}[0]{\begin{exampleHLP}\rm}{\end{exampleHLP}}
\newtheorem{remarkHLP}{Remark}[section]

\def\ema{{\mathcal M}_{\alpha}}
\def\dema{{\mathcal D}_{\alpha}}
\def\<{\langle}
\def\>{\rangle}
\begin{document}

\title{Affine connections, duality and divergences  for a von Neumann
algebra.}
\author{Anna Jen\v cov\'a}
\maketitle

\begin{center}{\Small Mathematical Institute, Slovak Academy of Sciences,\\
\v Stef\'anikova
49, 814 73 Bratislava, Slovakia,\\ jenca@mat.savba.sk}
\end{center}

\noindent {\bf Abstract.}
{\small On the predual of a von Neumann algebra,
we define a differentiable manifold structure and affine connections
by  embeddings into
non-commutative $L_p$--spaces. Using the geometry of uniformly convex Banach
spaces and  duality of the $L_p$ and $L_q$ spaces for $1/p+1/q=1$, we show that
we can introduce the $\alpha$-divergence, for $\alpha\in (-1,1)$, in a similar
manner as Amari in the classical case. If restricted to the positive  cone,
the $\alpha$-divergence belongs to the class of quasi-entropies, defined by
Petz.  }

\section{Introduction}

The classical information geometry deals with
the differential geometric aspects of families of probability densities with
respect to a given measure $\mu$.
The theory, developed in \cite{amari,chentsov}, has been
 already extended to the nonparametric
 case, where the manifold is modelled on some infinite dimensional
 Banach space, see \cite{pise,gipi}.

One of the important results of Amari's classical (finite dimensional)
information geometry \cite{amari,amana}
deals with the structure of Riemannian manifolds with a pair of flat affine
connections,
dual with respect to the metric.
For such manifolds, there is a pair $(\theta,\eta)$
of dual affine coordinate systems, related by  Legendre transformations
$$
\theta_i=\frac{\partial}{\partial\eta_i}\varphi(\eta)\quad 
\eta_i=\frac{\partial}{\partial\theta_i}\psi(\theta),
$$
where $\psi$, $\varphi$ are potential functions. A
quasi--distance, called the divergence, is then defined by
$$
D(\theta_1,\theta_2)=\psi(\theta_1)+\varphi(\eta_2)-\sum_i\theta_{1i}\eta_{2i}
$$
For manifolds of probability density functions, flat with respect to the $\pm
\alpha$--connections, the
corresponding $\alpha$-divergence belongs to the
class of Czisz\'ar's $f$-divergences
$$
S_f(p,q)=\int f(\frac qp)dp
$$
where $f$ is a convex function.
The $f$-divergences were generalized to  von Neumann algebras by Petz
in \cite{petz85} by means of the relative modular operator of normal positive
functionals on $M$:
$$
S_g(\phi,\psi)=(g(\Delta_{\phi,\psi})\xi_{\psi}, \xi_{\psi})
$$
where $\xi_{\psi}$ is the vector representative of $\psi$. On the other hand,
Amari's construction of the $\alpha$-divergence, starting from  a pair of dual 
flat
connections, was extended to the manifold of faithful positive linear
functionals  on a matrix algebra ${\mathcal M}_n(\mathbb {C})$, \cite{ja,
grasselli}. The aim of the present paper is to show that there is such a
construction for a general von Neumann algebra.

For $\alpha\in (-1,1)$, the $\alpha$-connections can be defined using
$\alpha$-embeddings into  non-commutative $L_p$-spaces, $p=\frac2{1-\alpha}$.
In this case, the  $\alpha$ and $-\alpha$-connections
are defined on different vector bundles and their duality
corresponds to the Banach space duality of $L_p$ and $L_q$, $1/p+1/q=1$,
therefore this duality does not require a Riemannian metric. This was shown  
by Gibilisco and Isola in \cite{giis} ( see also \cite{gipi} for the classical
case). Here, the $\alpha$-embeddings were used to define the
$\alpha$-connections on manifolds of faithful density operators of a 
semifinite von
Neumann algebra. The manifold structure, however, was not specified here, 
although some definitions of such a structure
already appeared, see \cite{grastre,streater00a,streater00b}.

Another possibility is to use the $\alpha$-embedding to introduce the manifold
structure. Here the problem is, that the range of the $\alpha$-embedding is in
the positive cone of the $L_p$-space which, even in the classical case, can
have empty interior. This problem was avoided in
\cite{ja03}, in defining the $\alpha$-embedding on the whole predual $M_*$ and
not just on the positive cone.

The $\alpha$-connections are defined as the trivial
connections in $L_p(M,\phi)$ and the $\pm \alpha$ -duality is just the Banach
space duality. The $\pm\alpha$-embeddings define a pair of dual coordinates on
$M_*$. Using the fact that the $L_p$ spaces with $p\in (1,\infty)$ are
uniformly convex, it was shown that the dual coordinates are related by
potential functionals, just as in Amari's theory. From this, we can define a
divergence functional on $L_p(M,\phi)$.

Via the $\alpha$-embedding, the divergence in $L_p(M,\phi)$ induces a
functional on $M_*\times M_*$, which is called the $\alpha$-divergence.
We will show that if restricted to the positive cone, the $\alpha$--divergence
is exactly the Petz
quasi-entropy $S_{g_{\alpha}}$, with
$$
g_{\alpha}(t):=\frac2{1-\alpha}+\frac2{1+\alpha}t-\frac4{1-\alpha^2}t^{\frac
{1+\alpha}2}.
$$

We will further investigate the properties of the divergence in $L_p(M,\phi)$,
especially the projection theorems. These imply some existence and uniqueness
results for the
$\alpha$--projections,
which generalize the projection theorems in \cite{amari}.

\section{Uniformly convex  Banach spaces.}\label{sec:banach}

We recall some facts about convexity and smoothness in Banach spaces, see
\cite{kothe}.

Let $X$ be a Banach space and let $X^*$ be the dual of $X$. 
Then for $u\in X^*$ we denote $\<x,u\>=u(x)$. Let
$K$ be a closed convex subset in $X$ with nonempty interior, in particular, let
$K_d$ be closed ball with radius $d$. Let $S$ be the boundary of $K$.

  A supporting hyperplane of $K$ is a real hyperplane $x+H$,
   containing at least
   one point of $K$ and such that $K$ lies in one of the two closed half-spaces
   determined by $x+H$.
   There is at least one supporting hyperplane through every boundary point of
   $K$.
   A boundary point $x_0\in S$ is called a  point of smoothness if exactly
   one closed supporting hyperplane passes through $x_0$, called a
   tangent hyperplane. We say that $K$ is smooth if every boundary point
   is a point of smoothness.
   The space $X$ is called smooth if $K_1$ is smooth.

     A normed space is smooth if and only if the norm is weakly
   differentiable at each point except the origin.
   The weak derivative of the norm at $x_0$ in the direction $y$ is
   given by $\Re\<y,v_{x_0/\|x_0\|}\>$, where $v_{x_0/\| x_0\|}$ is the unique
   point in the unit sphere of $X^*$, satisfying 
   $\<x_0,v_{x_0/\|x_0\|}\>=\| x_0\|$ and $\Re$ denotes the real part.
   The tangent hyperplane to the sphere $S_{\|x_0\|}$ 
   at $x_0$ is $x_0+H$, with
   $$
   H=\{ x\in X\ , \Re \<x, v_{x_0/\|x_0\|}\>=0\}
   $$

  The set $K$ is said to be  strictly convex if 
  every boundary point of $K$ is an extreme point,
  equivalently, the boundary of $K$  contains no line segment. In this case,
  each supporting hyperplane meets $K$ in exactly one point.

  A reflexive Banach space is smooth if and only if its dual $X^*$ is strictly
  convex, that is, the unit ball in $X^*$ is strictly convex. 

The space $X$ and its closed unit ball, are said to be uniformly convex if
for each $\epsilon$, $0<\epsilon\le 2$ there is a $\delta(\epsilon)>0$ such
that
$\| x\|\le 1$, $\|y\|\le 1$ and $\|x-y\|\ge \epsilon$ always implies that
$\|\frac12 (x+y)\|\le 1-\delta(\epsilon)$.
The function $\delta(\epsilon)$ is called the  module of convexity.
Every uniformly convex space is strictly convex
and  reflexive.

There is also a stronger notion of smoothness, dual to uniform convexity. The
space $X$, and its norm, are said to be uniformly smooth if for each
$\epsilon>0$ there is an $\eta(\epsilon)>0$, such that $\| x\|\ge 1$, $\|y\|\ge
1$ and $\|x-y\|\le \eta(\epsilon)$ always implies
$\|x+y\|\ge \|x\| +\|y\|-\epsilon\|x-y\|$.

A normed space $X$ is uniformly smooth if and only if its norm is uniformly
strongly differentiable. In particular, every uniformly smooth normed space is
smooth.
A Banach space $X$ is uniformly convex (uniformly smooth) if and
only if  $X^*$ is uniformly smooth (uniformly convex).

We will also need the following two results by Cudia \cite{cudia}.

\begin{theorem}\label{theorem:uniformlyctns}
 Let $S$ resp. $ S'$ be the unit sphere in $X$ resp. $X^*$. The norm is
 (uniformly) strongly differentiable in $S$ if and only if the map
 $v:x\mapsto v_x$ is single valued and (uniformly) continuous from the norm
 topology on $S$ to the norm topology on $S'$.
\end{theorem}

Let us now define the map $F:X\to X^*$ by
$$
F(x)=\left\{\begin{array}{lr}\|x\|v_{x/\|x\|}, & x\neq 0\\
                           0              , & x=0
			                              \end{array}\right.
$$

\begin{theorem}\label{theorem:homeomorphism} Let the Banach space $X$ be
uniformly convex and let the norm be strongly differentiable.
Then $F$ is a homeomorphism of $X$ onto $X^*$ (in the norm topologies).
\end{theorem}

\section{Non-commutative $L_p$-spaces.}\label{sec:lpspaces}

Let $M$ be a von Neumann algebra and let $\phi$ be a faithful normal
semifinite weight. We denote $N_{\phi}$ the set of $y\in M$ satisfying
$\phi(y^*y)<\infty$ and $M_0$ the set of all elements in $N_{\phi}\cap
N_{\phi}^*$, entire analytic with  respect to the modular automorphism
$\sigma^{\phi}_t$ associated with $\phi$. We also denote the GNS map by
$N_{\phi}\ni y\mapsto \eta_{\phi}(y)\in H_{\phi}$.

Let  $1\le p\le \infty$ and let $L_p(M,\phi)$ be the non-commutative $L_p$
space with respect to $\phi$, as defined by Araki and Masuda in
\cite{arma82,masuda83}. The elements of $L_p(M,\phi)$ are
closed operators acting on the Hilbert space $H_{\phi}$, satisfying
$$
TJ_{\phi}\sigma^{\phi}_{-i/p}(y)J_{\phi}\supset J_{\phi}yJ_{\phi}T,
$$
for all $y\in M_0$, such that the $L_p$-norm
$$
\| T\|_{p,\phi}=\{ \sup_{x\in M_0,\|x\|\le1}\|
|T|^{p/2}\eta_{\phi}(x)\|\}^{2/p}$$
is finite. Then $L_p(M,\phi)$ with the $L_p$-norm is a Banach space.
Let $1<p<\infty$, then $L_p(M,\phi)$  is uniformly convex and uniformly 
strongly differentiable. The dual space $L_p^*(M,\phi)$ is $L_q(M,\phi)$, with 
$1/p+1/q=1$, where the duality is given by
\begin{equation}\label{eq:lp_pairing}
\< T,T'\>_{\phi}=\lim_{y\to1}(\,T\eta_{\phi}(y)\,,T'\eta_{\phi}(y)\,)
\end{equation}
where $T\in L_p(M,\phi)$, $T'\in L_q(M,\phi)$. The limit is taken in the
*-strong topology with restriction $y\in M_0$, $\|y\|\le 1$.

Each $T\in L_p(M,\phi)$, $1\le p<\infty$, 
has a unique polar decomposition of the form
$$
T=u\Delta^{1/p}_{\psi,\phi}
$$
where $\psi\in M_*^+$, $u\in M$ is a partial isometry, such that the support
projection $s(\phi)=u^*u$ and $\Delta_{\psi,\phi}$ is the relative modular
operator, see Appendix C in \cite{arma82} for definition and basic properties.
On the other hand, each operator of this form is in $L_p(M,\phi)$.
The positive cone $L_p^+(M,\phi)$ is the set of positive operators in
$L_p(M,\phi)$ and we have
$$
L_p^+(M,\phi)=\{\Delta^{1/p}_{\psi,\phi},\ \psi\in M_*^+\}
$$

The identity
\begin{equation}\label{eq:lp_l1isom}
\varphi(au)=\<u\Delta_{\varphi,\phi},a^*\>_{\phi}
\end{equation}
for $a\in M$ 
gives an isometric isomorphism of $M_*$ and $L_1(M,\phi)$. Similarly,
$L_2(M,\phi)$ is isomorphic to $H_{\phi}$ by
$$
u\Delta^{1/2}_{\varphi,\phi}\mapsto u\xi_{\varphi},
$$
where $\xi_{\varphi}$ is the vector representative of $\varphi$ in the neutral
positive cone in $H_{\phi}$.

If $\tilde{\phi}$ is a different n.s.f. weight, then there is an isometric
isomorphism $\tau_p(\tilde{\phi},\phi):\ L_p(M,\phi)\to L_p(M,\tilde{\phi})$
and
\begin{equation}\label{eq:lp_tau}
\<T,T'\>_{\phi}=\<\tau_p(\tilde{\phi},\phi)T,
\tau_q(\tilde{\phi},\phi)T'\>_{\tilde{\phi}}
\end{equation}
holds for all $T\in L_p(M,\phi)$ and $T'\in L_q(M,\phi)$.

A bilinear form on  $L_p(M,\phi)\times L_q(M,\phi)$ is defined by 
$$
[T,T']_{\phi}=\<T,T^{'*}\>_{\phi},\quad T\in L_p(M,\phi),\ T'\in L_q(M,\phi)
$$
If $T_k\in L_{p_k}(M,\phi)$, $\sum_k 1/p_k=1/r$, then the product
$T=T_1...T_n$ is well defined as an element of $L_r(M,\phi)$ and
$$
\|T\|_r\le \|T_1\|_{p_1}\dots\|T_n\|_{p_n}
$$
If $r=1$, then
\begin{eqnarray}\label{eq:lp_bilin1}
[T_1\dots
T_n]_{\phi}&:=&[T,1]_{\phi}=[T_1\dots T_k,T_{k+1}\dots T_n]_{\phi}=\\ \nonumber
&=&[T_{k+1}\dots T_nT_1\dots T_k]_{\phi}
\end{eqnarray}
for each $1\le k\le n-1$ and
\begin{equation}\label{eq:lp_bilin2}
|[T_1\dots T_n]_{\phi}|\le \|T_1\|_{p_1}\dots\|T_n\|_{p_n}
\end{equation}

\section{ The $\alpha$-embeddings and affine connections}

Let $M$ be a von Neumann algebra and let $\phi$ be a faithful normal semifinite
weight. 

For $\-1< \alpha<1$, we define the non-commutative $\alpha$-embedding by
\begin{eqnarray*}
\ell_{\alpha}^{\phi}:\ M_*&\to& L_p(M,\phi), \quad p=\frac2{1-\alpha}\\
 \omega &\mapsto& pu\Delta_{\varphi,\phi}^{1/p}
\end{eqnarray*}
where $\omega(a)=\varphi(au),\ a\in M$ is the polar decomposition of $\omega$.
It is clear from
uniqueness of the polar decompositions that
$\ell^{\phi}_{\alpha}$ is bijective.
Moreover, it maps the hermitian (that is, $\omega(a^*)=\overline{\omega(a)}$ )
elements in $M_*$ onto the real Banach space
$L^h_p(M,\phi)$ of self-adjoint operators in the $L_p$-space and
$M_*^+$ onto the positive cone $L_p^+(M,\phi)$. 

If $\psi$ is a different f.n.s. weight, then the space
$L_p(M,\psi)$ is identified with $L_p(M,\phi)$ by the isometric
isomorphism $\tau_p(\psi,\phi)$. 
The corresponding $\alpha$-embeddings are related by
$$
\ell_{\alpha}^{\psi}=
\tau_p(\psi,\phi)\ell_{\alpha}^{\phi}
$$

We denote by $\ema$ the set $M_*$ with
the manifold structure induced from $\ell_{\alpha}^{\phi}$. Due to the above
isomorphism, 
the manifold structure does not depend from the choice of $\phi$.
For $\omega\in M_*$,  
$\ell_{\alpha}^{\phi}(\omega)\in L_p(M,\phi)$ will be called the 
$\alpha$-coordinate
of $\omega$.  The $-\alpha$-coordinate is an element of the
dual space $L_q(M,\phi)$, $1/p+1/q=1$. Moreover, for $\omega_1,\omega_2\in
M_*$ and a n.s.f. weight $\psi$, we have by (\ref{eq:lp_tau})
\begin{eqnarray}\label{eq:not_depend_on_phi}
\<\ell_{\alpha}^{\psi}(\omega_1),\ell_{-\alpha}^{\psi}(\omega_2)\>_{\psi}&=&
\<\tau_p(\psi,\phi)\ell_{\alpha}^{\phi}(\omega_1),\tau_q(\psi,\phi)
\ell_{-\alpha}^{\phi}(\omega_2)\>_{\psi}=\\ \nonumber
&=&\<\ell_{\alpha}^{\phi}(\omega_1),
\ell_{-\alpha}^{\phi}(\omega_2)\>_{\phi}
\end{eqnarray}
In the sequel, we will just 
write $\ell_{\alpha}$ instead of $\ell_{\alpha}^{\phi}$.
We will say that $\ell_{\alpha}(\omega)$ and
$\ell_{-\alpha}(\omega)\in L_q(M,\phi)$ are  dual coordinates of 
$\omega\in M_*$.

The trivial connection in $L_p(M,\phi)$ induces a globally flat affine
connection on the tangent bundle $T\ema$, called the $\alpha$-connection.
Let us recall
 that there is a one-to-one correspondence between affine
 connections and parallel transports on $T\ema$. If the connection
 is globally flat, the parallel transport is given by a family of
 isomorphisms $U_{x,y}:\ T_x(\ema)\to
 T_y(\ema)$, $x,y\in \ema$, satisfying
 \begin{enumerate}
 \item[(i)] $U_{x,x}=Id$,
 \item[(ii)] $U_{y,z}U_{x,y}=U_{x,z}$
 \end{enumerate}
 In our case, the tangent space $T_x(\ema)$ can be identified with
 $L_p(M,\phi)$ and the map $U_{x,y}$ is the identity map for all
$x,y\in \ema$.
 We define the dual connection as in \cite{gipi}, that is, a linear connection
 on the cotangent bundle $T^*\ema$, such that the corresponding
 parallel transport $U^*$ satisfies
 $$
 \<v,U_{x,y}^*(w)\>_{\phi}=\<U_{y,x}(v),w\>_{\phi}
 =\<v,w\>_{\phi}
 $$
 for $w\in (T_x(\ema))^*\equiv L_q(M,\phi)$ and $v\in
 T_y(\ema)$. Obviously, $U^*$ is the trivial parallel transport in
 $L_q(M,\phi)$,
 hence the dual of the $\alpha$-connection is the $-\alpha$-connection.

\section{Duality.}

Let $\omega\in M_*$. We will show how $\omega$  is
related to its dual coordinates.
\begin{prop} \label{prop:duality}
Let  $\omega\in M_*$, $\omega(a)=\psi(au)$
be the polar decomposition and let
$\psi_u(a)=\psi(u^*au)$. Then
$$
pq\psi_u(a)=\<\ell_{\alpha}(\omega),a^*\ell_{-\alpha}(\omega)\>_{\phi},\quad
a\in M.
$$
\end{prop}

\begin{proof}
We have from (\ref{eq:lp_l1isom}) and (\ref{eq:lp_bilin1}) that
\begin{eqnarray*}
\psi_u(a)&=&\<\Delta_{\psi,\phi},u^*a^*u\>_{\phi}=[\Delta_{\psi,\phi}u^*au]_{\phi}=[\Delta_{\psi,\phi}^{1/p}\Delta^{1/q}_{\psi,\phi}u^*au]_{\phi}=\\
&=&[u\Delta_{\psi,\phi}^{1/p},\Delta^{1/q}_{\psi,\phi}u^*a]_{\phi}=\frac
1{pq}\<\ell_{\alpha}(\omega),a^*\ell_{-\alpha}(\omega)\>_{\phi}
\end{eqnarray*}
\end{proof}

The $L_p$ spaces for $1<p<\infty$ are uniformly convex and uniformly smooth,
therefore we can use the results of Section \ref{sec:banach}.

The  map which sends the $\alpha$-coordinate $x=\ell_{\alpha}(\omega)$ of 
$\omega$ onto the  dual coordinate:
$$
x\mapsto \tilde x := \ell_{-\alpha}\ell_{\alpha}^{-1}(x)
$$
is called the duality map.
It is easy to see that 
for $x\in L_p(M,\phi)$ we have
\begin{equation}\label{eq:vxtilde}
v_{x/{\|x\|_p}}=\|\frac xp\|^{1-p}_p\frac {\tilde x}q
\end{equation}
and $\tilde x$ is the unique element in
$L_q(M,\phi)$, such that 
\begin{equation}\label{eq:norms}
\|\frac{\tilde x}q\|^q_q=\|\frac xp\|^p_p\qquad\mbox{and}\qquad  \Re \<x,\tilde
x\>_{\phi}=pq\|\frac xp\|^p_p.
\end{equation}
\begin{prop}\label{prop:homeomorphism}
The duality map is a homeomorphism $L_p(M,\phi)\to L_q(M,\phi)$.
\end{prop}

\begin{proof}
Clearly, $pu\Delta^{1/p}_{\psi,\phi}\mapsto qu\Delta^{1/q}_{\psi,\phi}$ is
continuous
at 0.
Further, let $F$ be the map defined
in Section \ref{sec:banach} and $x\ne 0$, then we have from
(\ref{eq:vxtilde})
$$
F(x)=\| x\|_p v_{x/\|x\|_p}=\frac{p^p}{pq}\| x\|_p^{2-p}\tilde x
$$
The statement  now follows from Theorem \ref{theorem:homeomorphism}.

\end{proof}

Let us define the function 
$\Psi_p:\ L_p(M,\phi)\to R^+$  by
$$
\Psi_p(x)=q\|\frac xp\|^p_p=q\varphi(1),
$$
where $x=pu\Delta^{1/p}_{\varphi,\phi}$. Then we have

\begin{prop}\label{prop:varphi}
$\Psi_p$ is strongly differentiable. The strong derivative at $x$
is given by 
$$
 D_y\Psi_p(x)=\Re\<y,\tilde x\>_{\phi},\quad y\in L_p(M,\phi)
$$
where $\tilde x$ is the dual coordinate. If $1/p+1/q=1$, then 
$$
\Psi_q(\tilde x)=\Re\<x,\tilde x\>_{\phi} - \Psi_p(x) 
$$

\end{prop}

\begin{proof} We have from the uniform smoothness of $L_p(M,\phi)$ that the
norm is strongly differentiable at all points except $x=0$ and 
$$
D_y\|x\|_p=\Re\< y,v_{x/\|x\|_p}\>_{\phi}
$$
It follows from (\ref{eq:vxtilde}) that for $x\ne 0$,
$$
D_y\Psi_p(x)=q\|\frac
xp\|^{p-1}_p\Re\<y,v_{x/\|x\|_p}\>_{\phi}=\Re\<y,\tilde x\>_{\phi} 
$$

As $p>1$, the function $\| \frac xp\|_p^p$ is strongly differentiable at
$x=0$ and
$$
D_y\Psi_p(0)=0=\Re \<y,\tilde 0\>_{\phi}
$$
The last equality is rather obvious.

\end{proof}

In the commutative  case, as well as on the manifold of positive definite
$n\times n$ matrices, $\Psi_p$ is the potential function in the
sense of Amari, see \cite{amari} and \cite{ja,grasselli}. In general, it is not
twice differentiable, but the above Proposition shows that the
Legendre transformations, relating the dual coordinate systems, are still 
valid.   
It will be also clear from the results of the next Section, that
$$
\Psi_q(\tilde x)=\sup_{y\in L_p(M,\phi)}
(\Re\<y,\tilde x\>_{\phi}-\Psi_p(y))
$$
hence $\Psi_q$ is the conjugate of the convex function $\Psi_p$.

\section{Divergence in $L_p(M,\phi)$.}\label{sec:diver}

Following \cite{amari}, the function
$D_p:\ L_p(M,\phi)\times L_p(M,\phi)\to R^+$, defined  by
$$
D_p(x,y)=\Psi_p(x)+\Psi_q(\tilde y)-\Re\<x,\tilde y\>_{\phi}
$$
is called the divergence. It has the following properties.

\begin{prop} \label{prop:Dprops}
\begin{enumerate}
\item[(i)] Let $f_p(t)=p+qt^p-pqt$. Then 
\begin{equation}\label{eq:Dpineq}
D_p(x,y)\ge \|\frac yp\|_p^pf_p(\frac{\|x\|_p}{\|y\|_p})
\end{equation}
for all $x,y\in L_p(M,\phi)$, where for $y=0$, we take the limit $\lim_{t\to0}
t^pf(s/t)=0$ for all $s$. In particular,
$D_p(x,y)\ge 0$ for all $x,y\in L_p(M,\phi)$ and equality is attained
if and only if $x=y$.
\item[(ii)] $D_p$ is jointly continuous and strongly differentiable in the
first variable.
\item[(iii)] $D_p(y,x)=D_q(\tilde x,\tilde y)$
\item[(iv)]  $D_p(x,y)+D_p(y,z)=D_p(x,z)+\Re\<x-y,\tilde z-\tilde y\>_{\phi}$
\end{enumerate}
\end{prop}

\begin{proof} The statement  (ii) follows from Proposition \ref{prop:varphi},
(iii) and (iv) follow easily from the definition of $D_p$.
We will now prove (i). If  $y=0$, then $D_p(x,y)=\Psi_p(x)\ge0$. Similarly,
if $x=0$, $D_p(x,y)=\Psi_q(\tilde y)$, which is equal to the right hand side
of (\ref{eq:Dpineq}).

 Let now $x\ne 0$, $y\ne 0$
and let $t=\|x\|_p/\|y\|_p$. Then by (\ref{eq:vxtilde})
$$
\Re\<x,\tilde y\>_{\phi}=tq\|\frac yp\|_p^{p-1}\Re \<\frac
xt,v_{y/\|y\|_p}\>_{\phi}
$$
Let $\|y\|_p=r$ and let $S_r$ be the sphere with radius $r$ in $L_p(M,\phi)$.
Then  $y, \frac xt\in S_r$.
From Section \ref{sec:banach}, the tangent hyperplane $y+H$ to
$S_r$ at $y$ is given by $\Re\<z,v_{y/r}\>_{\phi} =r$,
$S_r$ lies entirely in the half-space given by
$\Re\<z,v_{y/r}\>_{\phi}\leq r$ and $y$ is the
unique point of $S_r$ contained in $y+H$. Hence,
$$
D_p(x,y)\geq \Psi_p(x)+\Psi_q(\tilde y)-tpq\|\frac yp\|_p^p=\|\frac
yp\|_p^pf_p(t)\geq 0,
$$
where equality is attained in the first inequality if and only if $\frac xt=y$,
and in the second inequality if and only if $t=1$.
\end{proof}

We will also need the following lemma.
\begin{lemma}\label{lemma:uyd}
Let $y\in L_p(M,\phi)$, $d>0$ and let
$$
U_{y,d}:=\{ x\in L_p(M,\phi),\ D_p(x,y)\le d\} 
$$
Then $U_{y,d}$ is weakly closed, convex and  contains no half-line.
\end{lemma}

\begin{proof}
It is easy to see that $D_p$ is convex in the first variable, therefore
the set $U_{y,d}$ is also convex. Next,
let $\{ x_{\lambda}\}$ be a net in $U_{y,d}$, converging weakly to some
$x\in L_p(M,\phi)$ (it is in fact sufficient to consider sequences).
Then $0\le D_p(x_{\lambda},y)\le d$ and
we may suppose that the
net $d_{\lambda}=D_p(x_{\lambda},y)$ has a limit in $[0,d]$, using a
subnet  if necessary.  We have
$$ 
\lim_{\lambda} d_{\lambda}=\Psi_q(\tilde y)+\lim_{\lambda}
\{q\|\frac {x_{\lambda}}p\|_p^p -\<x_{\lambda},\tilde y\>_{\phi}\}.
$$ 
It follows that $\lim_{\lambda}\|x_{\lambda}\|_p$ exists.
Furthermore, for $u$ in the unit sphere of $L_q(M,\phi)$,
$$
|\<x,u\>_{\phi}|=\lim_{\lambda} |\<x_{\lambda},u\>_{\phi}|\le
\lim_{\lambda}\|x_{\lambda}\|_p
$$
and hence $\|x\|_p\le  \lim_{\lambda}\|x_{\lambda}\|_p$.  We therefore have
$$
D_p(x,y)=\Psi_q(\tilde y) + q\|\frac xp\|_p^p-\<x,\tilde y\>_{\phi}\le
\lim_{\lambda} d_{\lambda} \le d
$$
and $U_{y,d}$ is weakly closed.

Finally, let $h\ne 0$ and
let $x_t=x+th$, $t\ge 0$ be a half-
line in $L_p(M,\phi)$. For $y=0$, we have $D_p(x_t,0)=q\|\frac {x_t}p\|^p_p$.
If $y\ne 0$,  then by Proposition \ref{prop:Dprops} (i),
$$
D_p(x_t,y)\ge \|\frac yp\|^pf_p(\frac{\|x_t\|}{\|y\|})
$$
In both cases, the right-hand side goes to infinity as $t\to \infty$.
Therefore $U_{y,d}$ can contain no half--line.

\end{proof}

\section{$D_p$-projections.}\label{sec:projections}

Let $C$ be a subset in $L_p(M,\phi)$, $y\in L_p(M,\phi)$.
If there is a point $x_m\in C$, such that
$$
D_p(x_m,y)=\min_{x\in C}D_p(x,y)
$$
then $x_m$ will be called a  $D_p$-projection of $y$ to $C$. In this
section, we prove some uniqueness and existence results for $D_p$-projections.

\begin{prop}\label{prop:uniqueproj} Let $C$ be a convex subset in
$L_p(M,\phi)$, $y\in L_p(M,\phi)$ and $x_m\in C$. The following are equivalent.
\begin{enumerate}
\item[(i)] $D_p(x_m,y)=\min_{x\in C}D_p(x,y)$\\
\item[(ii)] $\tilde y-\tilde x_m$ is in the normal cone to $C$ at $x_m$,
that is,
$$
\Re \<x-x_m,\tilde y-\tilde x_m\>_{\phi}\le 0,\quad \forall x\in C
$$
\item[(iii)] $D_p(x,y)\ge D_p(x,x_m)+D_p(x_m,y), \quad \forall x\in C$

\end{enumerate} 
If such a point exists, it is unique.
\end{prop}

\begin{proof} Let $x_m$ be a point in $C$ satisfying (i) and let $x\in C$. Then
$x_t=tx+(1-t)x_m$ lies in $C$ for all $t\in [0,1]$ and thus 
$D_p(x_t,y)\ge D_p(x_m,y)$ on $[0,1]$. We have from Proposition \ref{prop:varphi}
$$
0\le \frac d{dt^+}D_p(x_t,y)|_{t=0}=\Re\<x-x_m,\tilde x_m-\tilde y\>_{\phi}
$$
which is (ii). Further, from Proposition \ref{prop:Dprops} (iv)
$$
\Re\<x-x_m,\tilde x_m-\tilde y\>_{\phi}=
D_p(x,y)-D_p(x,x_m)-D_p(x_m,y),
$$
hence (ii) implies (iii). 
Finally, let $x_m$ satisfy (iii), then we clearly have $D_p(x_m,y)\le
D_p(x,y)$, for all $x\in C$.

To prove uniqueness, suppose that $x_1$ and $x_2$ are points in $C$, satisfying
(iii). Then
$$
D_p(x_1,y)\ge D_p(x_1,x_2)+D_p(x_2,y)\ge D_p(x_1,x_2)+D_p(x_2,x_1)+D_p(x_1,y).
$$
It follows that $D_p(x_1,x_2)+D_p(x_2,x_1)\le 0$ and hence $x_1=x_2$.
\end{proof}

\begin{prop} Let $C$ be a weakly compact subset in $L_p(M,\phi)$ and
$y\in L_p(M,\phi)$. Then there exists a $D_p$-projection of $y$ to $C$.
\end{prop}

\begin{proof}
For some $d>0$, the  set $U_{y,d}$ has a nonempty intersection with $C$. 
By   Lemma \ref{lemma:uyd}, the sets $U_{y,d}\cap C$ are 
weakly closed. The intersection of these sets for all such $d$
is therefore nonempty
and is equal to some $U_{y,\rho}\cap C$. Then $\rho=\min_{x\in C}D_p(x,y)$ and
all the points in
$U_{y,\rho}\cap C$ are $D_p$-projections of $y$ in $C$.
\end{proof}

\begin{prop}\label{prop:existence}
Let $C$ be a weakly closed, convex, weakly locally compact subset in
$L_p(M,\phi)$. Then for each $y\in L_p(M,\phi)$ there  is a unique
$D_p$-projection to $C$.
\end{prop}

\begin{proof} Similarly as in the proof of previous Proposition, 
the set $U_{y,d}\cap C$ is  non-empty for sufficiently large  $d> 0$.
 By Lemma \ref{lemma:uyd}, this set
is convex and weakly closed.
As $C$ is weakly locally compact, $U_{y,d}\cap C$ is also
weakly locally compact. By \cite{kothe}, pp. 340, a closed convex locally
compact subset in a locally convex space is compact if and only if it contains
no half-line. It follows that $U_{y,d}\cap C$ are weakly compact and the
intersection of all such nonempty sets is therefore nonempty. Each point in
this intersection is a $D_p$-projection of $y$ to $C$.  By Proposition
\ref{prop:uniqueproj}, such a point is unique.
\end{proof}

Under the hypotheses of the above Proposition, we can define the map
$y\mapsto x_m$, which sends each point $y$ to its unique
$D_p$-projection in $C$.

\begin{prop} Let $C$ be a weakly closed convex weakly locally compact subset in
$L_p(M,\phi)$ and let $0\in C$.  
Then the $D_p$-projection is  continuous
from $L_p(M,\phi)$ with its norm topology to $C$ with the relative weak
topology.
\end{prop}

\begin{proof} Let $\{y^n\}$ be a sequence in $L_p(M,\phi)$
converging in norm to  $y$.
Let $x_m^n$ be the unique  $D_p$-projection of $y^n$  and
$x_m$ be the unique $D_p$-projection  of $y$ in $C$ from Proposition
\ref{prop:existence}. We have to prove that $x_m^n$ converges weakly to $x_m$.

Let $k>0$ be such that $\|y^n\|_p\le k$ for all $n$. 
Inserting $x=0$ in Proposition 
(\ref{prop:Dprops}), we get 
$$
0\le D_p(x_m,y)\le \Psi_q(\tilde y)-\Psi_q(\tilde x_m)
$$
and therefore by (\ref{eq:norms}),
$\|x_m\|_p\le \|y\|_p\le k$.  Similarly,  $\|x_m^n\|_p\le\|y^n\|_p\le  k$ for
each $n$.

As the duality map is continuous, we have $\tilde y^n\to \tilde y$ in
$L_q(M,\phi)$. Further, we have from joint continuity of $D_p$ that
$\lim D_p(y,y^n)=\lim D_p(y^n,y)=D_p(y,y)=0$. For sufficiently large $n$,
\begin{eqnarray*}
d_n&:=&D_p(x_m^n,y^n)=\inf_{x\in C,\|x\|_p\le
k}D_p(x,y^n)=\\ &=&\inf_{x\in C,\|x\|_p\le k}\{D_p(x,y)+D_p(y,y^n)-
\Re\<x-y, \tilde y^n-\tilde y\>_{\phi}\}\le\\ &\le & D_p(x_m,y)+D_p(y,y^n)+
2k\|\tilde y-\tilde y^n\|_q\le d+\varepsilon
\end{eqnarray*}
where $d:=D_p(x_m,y)$. Further,
\begin{eqnarray*}
D_p(x_m^n,y)&=&D_p(x_m^n,y^n)+D_p(y^n,y)-\Re\<x_m^n-y^n,\tilde y-\tilde
y^n\>_{\phi}\le\\ &\le & d_n + D_p(y^n,y)+2k\|\tilde y-\tilde y^n\|_q\le d+
2\varepsilon
\end{eqnarray*}
Hence for sufficiently large $n$, $x_m^n\in U_{y,d+2\varepsilon}\cap C$. As in
the proof of Proposition \ref{prop:existence}, these sets are nonempty weakly
compact sets and therefore $\{x_m^n\}$ contains a weakly convergent
subsequence. On the other hand, any limit of such subsequence has to be in
$U_{y,d+2\varepsilon}\cap C$ for all $\varepsilon$ and thus also in
$\bigcap_{\varepsilon} U_{y,d+2\varepsilon}\cap C$. This intersection contains
a single point $x_m$, it follows that $x_m^n$ converges weakly to $x_m$.
\end{proof}

\section{The $\alpha$-divergence in  $M_*^+$}

Let $\alpha\in (-1,1)$ and let $p=\frac 2{1-\alpha}$. 
The divergence in $L_p(M,\phi)$,  defines the functional
$S_{\alpha}:\ M_*\times M_*\to R^+$, by
\begin{eqnarray*}
S_{\alpha}(\omega_1,\omega_2)&:=&
D_p(\ell_{\alpha}(\omega_1),\ell_{\alpha}(\omega_2))=\\
&=&q\varphi(1)+p\psi(1)-pq\Re\<u\Delta^{1/p}_{\varphi,\phi},
v\Delta^{1/q}_{\psi,\phi}\>_{\phi}
\end{eqnarray*}
where $\omega_1(a)=\varphi(au)$ and $\omega_2(a)=\psi(av)$ are the polar
decompositions. It is called the $\alpha$-divergence. 
It follows from (\ref{eq:not_depend_on_phi}) that
$S_{\alpha}$ does not depend from $\phi$. In particular, 
if  $\psi$ is faithful, then 
$$
\<u\Delta^{1/p}_{\varphi,\phi},
v\Delta^{1/q}_{\psi,\phi}\>_{\phi}=(\Delta_{\varphi,\xi_{\psi}}^{1/(2p)}
\xi_{\psi},\Delta_{\varphi,\xi_{\psi}}^{1/(2p)}u^*v\xi_{\psi})
$$
where $\xi_{\psi}$ is a vector representative of $\psi$. 
It follows that if $\varphi,\psi\in M_*^+$, $\psi$  is
faithful and $\Delta_{\varphi,x_{\psi}}=\int \lambda E_{\lambda}$ is the
spectral decomposition, then
$$
S_{\alpha}(\varphi,\psi)=
(g_p(\Delta_{\varphi,\xi_{\psi}})\xi_{\psi},\xi_{\psi})=\int g_p(\lambda)
\|E_{\lambda}\xi_{\psi}\|^2
$$
where 
$
g_p(t)=p+qt-pqt^{1/p}$.
Hence, in this case the $\alpha$-divergence
is equal to the quasi entropy $S^1_{g_p}$, defined by Petz in
\cite{petz85,ohypetz93}. 
We will show that this is true on the whole of $M_*^+\times M_*^+$.

\begin{lemma}
Let $\varphi,\psi\in M_*^+$, $u,v\in M$ be partial isometries satisfying
$u^*u=s(\varphi)$, $v^*v=s(\psi)$. Let $p,q>1$ be such that
$\frac1p+\frac1q=1$. Then
\begin{equation}\label{eq:same}
\<u\Delta_{\varphi,\phi}^{1/p},v\Delta_{\psi,\phi}^{1/q}\>_{\phi}= 
(\,\Delta_{\varphi,\xi_{\psi}}^{1/(2p)}\xi_{\psi},\Delta_{\varphi,\xi_{\psi}}^{1/(2p)}u^*v\xi_{\psi}\,)
\end{equation}
where $\xi_{\psi}$ is a vector representative of $\psi$.
\end{lemma}
\begin{proof} Let $1/p\le 1/2$. We have
$$
\<u\Delta_{\varphi,\phi}^{1/p}\,,v\Delta_{\psi,\phi}^{1/q}\>_{\phi}=
\lim_{y\to 1} 
(\,
\Delta^{1/2-1/p}_{\psi,\phi}v^*u\Delta_{\varphi,\phi}^{1/p}\eta_{\phi}(y)\,,\Delta_{\psi,\phi}^{1/2}\eta_{\phi}(y)\,),
$$
with $y\in M_0$, $\|y\|\le 1$. For $y\in N_{\phi}$,
\begin{eqnarray}\label{eq:pomoc}
&(\,\Delta^{1/2-1/p}_{\psi,\phi}v^*u\Delta_{\varphi,\phi}^{1/p}\eta_{\phi}(y)
\, ,
\Delta_{\psi,\phi}^{1/2}\eta_{\phi}(y)\,)=\\
&=
(\,J_{\xi_{\psi},\eta_{\phi}}\Delta_{\psi,\phi}^{1/2}\eta_{\phi}(y)\,,
J_{\xi_{\psi},\eta_{\phi}}\Delta_{\psi,\phi}^{1/2-1/p}v^*u
\Delta_{\varphi,\phi}^{1/p}\eta_{\phi}(y)\,)=\nonumber\\
&=(\,y^*\xi_{\psi}\,,J_{\xi_{\psi},\eta_{\phi}}\Delta_{\psi,\phi}^{1/2-1/p}v^*u
\Delta_{\varphi,\phi}^{1/p}\eta_{\phi}(y)\,),\nonumber
\end{eqnarray}
here we have used that $J_{\xi_{\psi},\eta_{\phi}}^*J_{\xi_{\psi},\eta_{\phi}}=
s(\psi)=s(\Delta_{\psi,\phi})$, the support of $\Delta_{\psi,\phi}$.
Let $t\in \mathbb{R}$, then
\begin{eqnarray*}
&J_{\xi_{\psi},\eta_{\phi}}\Delta_{\psi,\phi}^{1/2-it}v^*u
\Delta_{\varphi,\phi}^{it}\eta_{\phi}(y)=
S_{\xi_{\psi},\eta_{\phi}}
\Delta_{\psi,\phi}^{-it}v^*u\Delta_{\varphi,\phi}^{it}\eta_{\phi}(y)=\\
&S_{\xi_{\psi},\eta_{\phi}}v^*
\Delta_{\psi_v,\phi}^{-it}\Delta_{\varphi_u,\phi}^{it}u\eta_{\phi}(y)=
S_{\xi_{\psi},\eta_{\phi}}v^*(D\psi_v:D\varphi_u)_{-t}u\eta_{\phi}(y)=\\
&y^*u^*(D\psi_v:D\varphi_u)^*_{-t}v\xi_{\psi}
\end{eqnarray*}
where $\varphi_u(a)=\varphi(u^*au)$ and
$u\Delta_{\varphi,\phi}^{it}u^*=\Delta_{\varphi_u,\phi}^{it}$ by (C.8) in
\cite{arma82}. From this, we have
\begin{eqnarray*}
(\,y^*\xi_{\psi}\,,J_{\xi_{\psi},\eta_{\phi}}\Delta_{\psi,\phi}^{1/2-it}v^*u
\Delta_{\varphi,\phi}^{it}\eta_{\phi}(y)\,)= 
(\,(D\psi_v:D\varphi_u)_{-t}uyy^*\xi_{\psi}\,, v\xi_{\psi}\,)=\\
=(\, \Delta_{\psi_v,\xi_{\psi}}^{-it}\Delta_{\varphi_u,\xi_{\psi}}^{it}uyy^*
\xi_{\psi}\,, v\xi_{\psi}\,)=(\,
u\Delta_{\varphi,\xi_{\psi}}^{it}yy^*\xi_{\psi}\,,v\xi_{\psi}\,),
\end{eqnarray*}
where we have used (C.5) and (C.8) of \cite{arma82}. It follows that for
$z=it$,
\begin{equation}\label{eq:holo}
(\,y^*\xi_{\psi}\,,J_{\xi_{\psi},\eta_{\phi}}\Delta_{\psi,\phi}^{1/2-z}v^*u
\Delta_{\varphi,\phi}^z\eta_{\phi}(y)\,)= (y^*\xi_{\psi}\,, y^*\Delta_{\varphi,
\xi_{\psi}}^{\bar z}u^*v\xi_{\psi}\,).
\end{equation}
By Lemma 3.1 in \cite{masuda83}, both sides of (\ref{eq:holo}) are holomorphic
for $0<\Re z <1/2$ and continuous for $0\le \Re z\le 1/2$. The 
equation
(\ref{eq:same}) holds  for $1/p\le1/2$ by (\ref{eq:pomoc}) and analytic continuation of
(\ref{eq:holo}).

Let now $1/q\le1/2$. We have by the first part of the proof
\begin{eqnarray*}
\<u\Delta_{\varphi,\phi}^{1/p}\,,v\Delta_{\psi,\phi}^{1/q}\>_{\phi}&=& (\,
u\xi_{\varphi}\,,v\Delta_{\xi_{\psi},\xi_{\varphi}}^{1/q}\xi_{\varphi}\, )=
(\,S_{\xi_{\varphi},\xi_{\psi}}u^*v\xi_{\psi}\,,
\Delta_{\xi_{\psi},\xi_{\varphi}}^{1/q}S_{\xi_{\varphi},\xi_{\psi}}\xi_{\psi}\,)\\&=&
(\,
J_{\xi_{\psi},\xi_{\varphi}}\Delta_{\xi_{\psi},\xi_{\varphi}}^{1/q}
J_{\xi_{\varphi},\xi_{\psi}}\Delta_{\xi_{\varphi},\xi_{\psi}}^{1/2}\xi_{\psi}\,,\Delta_{\xi_{\varphi},\xi_{\psi}}^{1/2}u^*v\xi_{\psi}\,)=\\
&=&(\, 
\Delta_{\varphi,\xi_{\psi}}^{1/p-1/2}\xi_{\psi},
\Delta^{1/2}_{\varphi,\xi_{\psi}}u^*v\xi_{\psi}\,),
\end{eqnarray*}
we have used the equations (C.14) $J_{\eta_1,\eta_2}^*=J_{\eta_2,\eta_1}$
and ($\beta$5)
$J_{\eta_1,\eta_2}\Delta_{\eta_1,\eta_2}J_{\eta_2,\eta_1}=
\Delta_{\eta_2,\eta_1}^{-1}$ from Appendix C in \cite{arma82}.
\end{proof}

It follows that
$S_{\alpha}(\varphi,\psi)=S^1_{g_p}(\varphi,\psi)$ for all positive normal
functionals $\varphi$ and $\psi$.
The function $g_p$, $1<p<\infty$ is operator convex and it follows from the
results  in \cite{petz85} that
\begin{enumerate}
\item[(i)] $S_{\alpha}$ is jointly convex on $M_*^+\times M_*^+$
\item[(ii)] $S_{\alpha}$ decreases under stochastic maps on $M_*^+\times M_*^+$
\item[(iii)]
$S_{\alpha}$ is lower semicontinuous on $M*^+\times {\mathcal F}(M_*^+)$
endowed with the product of norm topologies, where ${\mathcal F} (M_*^+)$
denotes the set of faithful elements in $M_*^+$.
\end{enumerate}

The following properties of the $\alpha$-divergence are valid on $M_*\times
M_*$ and  are immediate consequences
of the results of Section \ref{sec:diver}.
\begin{enumerate}
\item[(i)] Positivity
$$
S_{\alpha}(\varphi,\psi)\ge \|\psi\|_1g_p(\frac{\|\varphi\|_1}{\|\psi\|_1})
\ge 0
$$
and $S_{\alpha}(\varphi,\psi)=0$ if and only if $\varphi=\psi$ (here
$\|\cdot\|_1$ is the norm in $M_*$).
\item[(ii)] $S_{\alpha}(\varphi,\psi)=S_{-\alpha}(\psi,\varphi)$
\item[(iii)] generalized Pythagorean relation
$$
S_{\alpha}(\varphi,\psi)+S_{\alpha}(\psi,\sigma)=S_{\alpha}(\varphi,\sigma)+
\Re\<\ell_{\alpha}(\varphi)-\ell_{\alpha}(\psi),\ell_{-\alpha}(\sigma)-
\ell_{-\alpha}(\psi)\>_{\phi}
$$
\end{enumerate}

Notice that the Pythagorean relation (iii) is a generalization of the classical
version in \cite{amari}, which says that equality is attained if and only if 
the $\alpha$-geodesic connecting $\psi$ and $\varphi$ is orthogonal to the
$-\alpha$-geodesic connecting $\psi$ and $\sigma$.

We also define the  $\alpha$-projection of $\varphi \in M_*$
onto a subset $C\subset M_*$ as the element in $C$ that minimizes 
$S_{\alpha}(\cdot,\varphi)$ over $C$. We will say that a subset $C\subset M_*$
is $\alpha$-convex if $\ell_{\alpha}(C)$ is  convex. The next Proposition is
a generalization of the results in \cite{amari,amana} and follows directly from
Proposition \ref{prop:uniqueproj}.

\begin{prop}
Let $C\subset M_*$ be $\alpha$-convex and let $\psi\in M_*$, $\varphi_m\in C$. 
The following are equivalent.
\begin{enumerate}
\item[(i)] $\varphi_m$ is an $\alpha$-projection of $\psi$ in $C$.
\item[(ii)] For all $\sigma\in C$,
$$
S_{\alpha}(\sigma,\psi)\ge
S_{\alpha}(\varphi_m,\psi)+S_{-\alpha}(\varphi_m,\sigma)
$$
\item[(iii)] 
The curve $x_t\in L_q(M,\phi)$, 
$$x_t:=\ell_{-\alpha}(\varphi_m) +
t(\ell_{-\alpha}(\psi)-\ell_{-\alpha}(\varphi_m))
$$ 
lies in the normal
cone to $\ell_{\alpha}(C)$ at $\ell_{\alpha}(\varphi_m)$ for all $t\ge 0$ 
(Note that
$\ell^{-1}_{-\alpha}(x_t)$ is the $-\alpha$-geodesic connecting $\varphi_m$ and
$\psi$.)
\end{enumerate}
If such a point exists, it is unique.
\end{prop}

The topology induced by the $\alpha$-embedding from the norm, resp. the weak
topology in $L_p(M,\phi)$ will be called the $\alpha$-, resp. the $\alpha$-weak
topology. The following Proposition is also immediate from Section
\ref{sec:projections}.

\begin{prop}\label{prop:alphaproj} 
Let $C\subset M_*$ and let $\psi\in M_*$.
\begin{enumerate}
\item[(i)] If $C$ is $\alpha$-weakly compact, 
then there exists an $\alpha$-projection of $\psi$ in $C$.
\item[(ii)] If $C$ is $\alpha$-weakly closed, $\alpha$-convex, $\alpha$-weakly
locally compact, then there exist a unique projection of $\psi$ in $C$.
\item[(iii)] If $C$ is as in (ii) and, moreover, $0\in C$, then the
$\alpha$-projection is a continuous map from $M_*$ with the $\alpha$-topology
to $C$ with the relative $\alpha$-weak topology.
\end{enumerate}

\end{prop}

\begin{example} Let $C$ be an extended $\alpha$-family, generated by a finite
number of positive elements, that is, there exist $x_1,\dots,x_n\in
L_p(M,\phi)$, such that
$$
\ell_{\alpha}(C)=\{\sum_{i=1}^nt_ix_i,\ t_i\geq 0, i=1,\dots,n\}
$$
It follows from Proposition \ref{prop:alphaproj} (iii) that we have an 
$\alpha$-projection from $M_*$ to $C$, which is continuous
in the $\alpha$-topology. 
\end{example}

\section{The case $\alpha=0$.}\label{sec:nula}

Let $\alpha=0$, $p=q=2$. The space $L_2(M,\phi)$ can be identified with the
Hilbert space $H_{\phi}$ and the dual
pairing $\<\cdot,\cdot\>_{\phi}$ is the inner product $(\cdot,\cdot)$ in
$H_{\phi}$.
Through this identification, the $0$-embedding becomes the map 
$$
\omega\mapsto 2u\xi_{\varphi}
$$
where $\omega(a)=\varphi(au)$ is the polar decomposition of 
$\omega$ and $\xi_{\varphi}$ is 
the unique vector representative of $\varphi$ in the neutral 
positive cone $V$ in $H_{\phi}$.
Hence the $0$-embedding maps $M_*$ bijectively onto $H_{\phi}$. In this case, 
the duality map is the identity on $H_{\phi}$ and the potential function is
$$
\varphi_2(x)=\frac 12\|x\|^2
$$
Therefore, the potential function is $C^{\infty}$-diferentiable and
$$
D^2_{y,z}\varphi_2(x)=\Re(y,z)\qquad  \forall x\in H_{\phi}
$$
It follows that $\varphi_2$ defines a Riemannian metric in the tangent
bundle $T{\mathcal M}_0$, which corresponds to the real part of the 
inner product, induced from the $0$-embedding. In the matrix case, this
metric was studied on density matrices  and it was shown
that it coincides with the Wigner--Yanase metric, see \cite{giis03}.

Up to multiplication by  2, the restriction of $\ell_0$ to the positive cone $M_*^+$ corresponds to
the identification of the positive normal functionals with elements in $V$
proved by Araki in
\cite{araki74}. It has been also shown that this identification
is a homeomorphism 
$M^+_*\to V$. It follows that the relative $0$-topology is the same as the
relative $L_1$-topology in $M_*^+$.

The $D_2$-divergence in $H_{\phi}$ is
$$
D_2(x,y)=\frac 12\|x-y\|^2,
$$
hence the $D_2$-projection corresponds to minimizing the Hilbert space norm.
This means, in particular, that there is a unique $D_2$-projection onto every 
closed convex subset of $H_{\phi}$.

The $0$-divergence in $M_*$ becomes
$$
S_0(\omega_1,\omega_2)=2\|u\xi_{\varphi}-v\xi_{\psi}\|^2
$$
On the positive cone, 
the $0$-divergence generalizes the classical Hellinger distance.

\section{Topologies induced in $M_*^+$}

In this section, we study various topologies induced by the
$\alpha$-embeddings in $M_*^+$. First of all, we see from Proposition
\ref{prop:homeomorphism} that the $+\alpha$- and $-\alpha$-topologies are the
same. Let now $\varphi,\psi\in M_*^+$ and let $\ell_{\alpha}(\varphi)=x$.
$\ell_{\alpha}(\psi)=y$. By Proposition \ref{prop:duality} and
(\ref{eq:lp_bilin2}), we have for $a\in
M$,
\begin{eqnarray*}
|\varphi(a)-\psi(a)|&=&|\<x,a^*\tilde x\>_{\phi}-\<y,a^*\tilde y\>_{\phi}|=\\
&=&\frac12 |\<(x+y),a^*(\tilde x-\tilde y)\>_{\phi}+\<(x-y), 
a^*(\tilde x+\tilde y)\>_{\phi}|\le\\
&\le& \frac 12\|a\|(\|x+y\|_p\|\tilde x-\tilde
y\|_q+\|x-y\|_p\|\tilde x+\tilde y\|_q)
\end{eqnarray*}
It follows that the map $\ell_{\alpha}^{-1}:\ L_p^+(M,\phi)\to M_*^+$ is
continuous relative to the norm topologies. Hence the $\alpha$-topology is
stronger than the $L_1$-topology in $M_*^+$.

Since the $\alpha$-divergences can be seen as quasi-distances in $M_*^+$, we
will consider the topology induced by $S_{\alpha}$, which will be called the 
$S_{\alpha}$-topology. The $S_{\alpha}$ topology is given by the base of
neighborhoods
$$
O^{\alpha}(\psi,\varepsilon):=\{\varphi\in M_*^+ ,\ S_{\alpha}(\varphi,\psi)<
\varepsilon\}
$$
for $\psi\in M_*^+$, $\varepsilon >0$. Because the functions $L_p(M,\phi)\ni
x\mapsto D_p(x,y)\in R^+$
are continuous for each $y$, the $S_{\alpha}$-topology is weaker than the
$\alpha$-topology. 

\begin{lemma}\label{lemma:alpha} 
Let $\varphi,\psi\in M_*^+$ and let $-1<\alpha\le\beta<1$. Then
\begin{eqnarray*}
(1-\beta)S_{\beta}(\varphi,\psi)&\le&
(1-\alpha)S_{\alpha}(\varphi,\psi)\\
(1+\alpha)S_{\alpha}(\varphi,\psi)&\le&
(1+\beta)S_{\beta}(\varphi,\psi)
\end{eqnarray*}
\end{lemma}

\begin{proof}
The proof is essentially the same as in the classical case, see for
example \cite{lievaj87}.

Let us consider the function
$$
F_t(a)=t^a-at+a-1\qquad a\in (0,1)
$$
Then $F_t$ is convex on $(0,1)$ for all $t\ge 0$. It follows that
$$
\frac{F_t(1)-F_t(a)}{1-a}\le\frac {F_t(1)-F_t(b)}{1-b}
$$
for all $0<a\le b<1$ and $t\ge0$. As $F_t(1)=0$ for all $t$, we get that the
function $\frac {F_t(a)}{a-1}$ is increasing on $(0,1)$. Let now $p=\frac
2{1-\alpha}$ and put $a=1/p$, then the function
$$
\frac {F_t(1/p)}{1/p-1}=1/pg_p(t)
$$
is decreasing on $(0,\infty)$. Hence we have for $0<p\le p'<\infty$
$$
\frac 1{p'}(g_{p'}(\Delta_{\varphi,\psi})\xi_{\psi},\xi_{\psi})\le   
\frac 1p(g_{p}(\Delta_{\varphi,\psi})\xi_{\psi},\xi_{\psi})
$$
and the first inequality follows. The second inequality is obtained from the
first and from $S_{\alpha}(\varphi,\psi)=S_{-\alpha}(\psi,\varphi)$.
\end{proof}

From the last Lemma, we get for $\varphi\in M_*^+$, $-1<\alpha\le\beta<1$ and
$d>0$,
$$
O^{\alpha}(\psi,\frac{1-\beta}{1-\alpha}d)\subseteq O^{\beta}(\psi,d)
\subseteq O^{\alpha}(\psi,\frac{1+\alpha}{1+\beta}d)
$$
hence the $S_{\alpha}$-topologies are the same for all $\alpha\in (-1,1)$. In
particular, these are the same as the $S_0$-topology, which, by Section
\ref{sec:nula}, is the same as the $0$-topology. It follows that on the
positive cone, the
topology induced from $S_{\alpha}$  coincides with the $L_1$-topology .

\section{The unit sphere.}

The $\alpha$-embedding maps the unit sphere $S$ 
in $M_*$ onto the sphere $S_p$ with
radius $p$ in $L_p(M,\phi)$.  The duality map $x\mapsto \tilde{x}$ maps $S_p$
onto the sphere $S'_q$ with radius $q$ in the dual space $L_q(M,\phi)$. 
From (\ref{eq:vxtilde}), we have that for $x\in S_p$,
\begin{equation}\label{eq:unit_vxtilde}
\tilde x = qv_{x/p}
\end{equation}

\begin{prop}The duality map $S_p\ni x\mapsto \tilde x \in S_q'$ is uniformly
continuous.
\end{prop}

\begin{proof} The statement 
follows from (\ref{eq:unit_vxtilde}) and Theorem \ref{theorem:uniformlyctns}.
\end{proof}

Further, there is a unique tangent hyperplane $x+H_x$ through $x$, where $H_x$
is given by the condition 
$$
\Re\<y,\tilde x\>_{\phi}=q\Re\<y,v_{x/p}\>_{\phi}=0
$$
Hence there is a splitting $L_p(M,\phi)=H_x\oplus [x]$ and, similarly as in
\cite{gipi}, there is a continuous projection $\pi_x:L_p(M,\phi)\to H_x$, given
by
$$
\pi_x(y)=y-\Re\<y,v_{x/\|x\|_p}\>_{\phi}\frac
xp=y-\frac1{pq}\Re\<y,\tilde x\>_{\phi}x,
$$
which is obtained by minimizing the $L_p$-norm.

As the norm is strongly differentiable, the unit
sphere can be given the structure of
a differentiable submanifold $\dema$ in $\ema$. If $\psi\in \dema$ has the
$\alpha$-coordinate $x\in S_p$, then the tangent space $T_x(\dema)$ can be
identified with the tangent hyperplane $H_x$ and  $\pi_x$ can be
used to project the $\alpha$-conection onto $T\dema$. But, even in the
classical and  the matrix case, the projected connection is no longer flat.
Hence, it does not define a divergence, but nevertheless, we can use the
restriction of $S_{\alpha}$ as a quasi-distance on $S$. This restriction has
the form

$$
S_{\alpha}(\omega_1,\omega_2)=pq(1-\Re\<u\Delta_{\varphi,\phi}^{1/p},v\Delta_{\psi,\phi}^{1/q}\>_{\phi})
$$
which corresponds to the definition of the $\alpha$-divergence in \cite{amari} 
for probability densities and in \cite{hase93} for density matrices.

Let us now consider the topologies induced on the set of states $S^+\subset
M_*^+$. 
From
 \cite{kothe} pp. 354, we have  that the weak and the strong topologies coincide
 on the unit sphere of a uniformly convex space, hence these coincide on $S_p$.
 It follows that the relative $\alpha$-topology and the $\alpha$-weak topology
 are the same on $S$.

 Let now $\varphi$, $\psi \in S$ and let $\ell_{\alpha}(\varphi)=x$,
 $\ell_{\alpha}(\psi)=y$. Then $x,y\in S_p\subset L_p(M,\phi)$ and
 $$
 \|\frac 12(\frac xp+\frac yp)\|_p\ge \frac 1{2pq}|\Re\<x+y,{\tilde
 y}\>_{\phi}|=|1-\frac 1{2pq}D_p(x,y)|
 $$
 Therefore if $D_p(x,y)<2pq\delta(\varepsilon)$, where
 $\delta(\varepsilon)$ is the module of convexity,
 then $\|\frac 12 (\frac xp+\frac yp)\|_p>1-\delta(\varepsilon)$
 and uniform convexity implies that $\|x-y\|_p<p\varepsilon$. It follows that
 for each $\varepsilon >0$,
 the set $S_p\cap \ell_{\alpha}(O^{\alpha}(\psi,2pq\delta(\varepsilon/p)))$ is contained in the strong
 neighborhood  $S_p\cap\|x-y\|_p<\varepsilon$. Therefore, the
 $S_{\alpha}$-topology coincides with the $\alpha$-topology on $S$. We have
 proved the following

 \begin{prop} The topologies on $S^+$, inherited from the $\alpha$-topology,
 $\alpha$-weak topology and $S_{\alpha}$-topology coincide with the
 $L_1$-topology for all $\alpha\in (-1,1)$.
 \end{prop}

\begin{coro}  The restriction of $S_{\alpha}$ to $S^+\times S^+$ is
continuous in the $L_1$-topology.
\end{coro}

\end{document}